\documentclass[twocolumn,showpacs,preprintnumbers,amsmath,amssymb,floatfix]{revtex4}
\usepackage{graphicx}

\begin{document}


\title{Nonlinear charge injection in organic field-effect transistors}

\author{B.H. Hamadani and D. Natelson}

\affiliation{Department of Physics and Astronomy, Rice University, 6100 Main St., Houston, TX 77005}

\date{\today}

\begin{abstract}

Transport properties of a series of poly(3-hexylthiophene) organic
field effect transistors with Cr, Cu and Au source/drain electrodes
were examined over a broad temperature range. The current-voltage
characteristics of the injecting contacts are extracted from the
dependence of conductance on channel length.  With reasonable
parameters, a model of hopping injection into a disordered density of
localized states, with emphasis on the primary injection event, agrees
well with the field and the temperature dependence of the
data over a broad range of temperatures and gate voltages.

\end{abstract}

\maketitle 


\section*{I. INTRODUCTION}

Field effect transistors based on organic semiconductors (OFETs) have
attracted interest for their potential applications as inexpensive and
flexible electronics\cite{DimitrakopoulosetAl01IBM}. The physics of
charge injection at the metal-organic semiconductor (OSC) contact in
such field-effect devices remains poorly understood. Contact
resistance is often neglected when inferring the mobility from
transistor transport characteristics.  It is known, however, that the
parasitic series resistance, $R_{\rm s}$, at the OSC-metal interface
can play an important role\cite{Scott03JVSTA}. Several experiments
have already shown that can easily dominate the intrinsic channel
resistance, $R_{\rm ch}$, in short channel (few microns and below)
devices\cite{KlauketAl03SSE,NecludiovetAl03SSE,HamadanietAl04APL}.
Approaches used to differentiate between contact and channel
resistances include analyses of single device
characteristics\cite{HorowitzetAl99JAP,StreetetAl02APL}, scanning
potentiometry\cite{SeshadrietAl01APL,BurgietAl02APL,BurgietAl03JAP},
and scaling of total device resistance with channel length in a series
of
devices\cite{KlauketAl03SSE,NecludiovetAl03SSE,HamadanietAl04APL,ZaumseiletAl03JAP,MeijeretAl03APL,BlanchetetAl04APL}.

In a previous study\cite{HamadanietAl04APL}, we reported measurements
of $R_{\rm s}$ and the true channel resistance in bottom contact poly
(3-hexylthiophene) (P3HT) field effect transistors as a function of
temperature. In that study, which used Au source and the drain
electrodes, we found that $R_{\rm s}$ correlates inversely with
mobility over four decades, over a broad range of temperatures and
gate voltages. This is consistent with the predictions of a recent
theory\cite{ScottetAl99CPL,ShenetAl01PRL} of OSC-metal contacts
incorporating a thermionic emission model with diffusion-limited
injection currents and accounting for the backflow of charge at the
interface.  Such a model predicts an inverse relationship between the
mobility and the contact resistivity, provided that the Schottky
barrier between Au and P3HT is low.  This is expected to be the case
since the highest occupied molecular orbital (HOMO) of P3HT is
estimated\cite{ChiguvareetAl03JAP} to lie between 5.1 and 5.2~eV,
close to the work function of Au (5.2~eV)\cite{BurgietAl03JAP}.

In the case of lower work function metals such as Cr and Cu ($\sim$
4.7~eV) \cite{BurgietAl03JAP}, a significant Schottky energy
barrier, $\Delta$, for holes is expected to exist at the OSC-metal
interface. For a channel mobility, $\mu$, that is thermally activated
with characteristic energy $E_{\rm a}$, typically $< \sim$0.1~eV, the
same model predicts a temperature dependence of the contact
resistance\cite{HamadanietAl04APL}, $R_{\rm s} \propto \exp{([E_{\rm
      a}+\Delta]/k_{\rm B}T)}$.  For $\Delta \sim$~0.3~eV, the
temperature dependence of $R_{\rm s}$ is therefore predicted to be
much stronger than the Au electrode case.  However, in recent studies
of charge injection in both bottom-contact P3HT field effect
transistors\cite{BurgietAl03JAP} and hole injection from an Ag
electrode into poly-dialkoxy-p-phenylene
vinylene\cite{WoudenberghetAl01APL}, only a weak temperature
dependence of contact resistance or injecting current was
observed.  These results imply that the diffusion-limited thermionic
emission model is inadequate.

In this work, we examine this issue through the temperature and field
dependence of charge injection in bottom contact OFETs based on P3HT
with different source/drain electrode materials.  To differentiate
between the current-voltage characteristics of the channel and the
contacts, we examine the scaling of device current with channel
length, employ the gradual channel
approximation\cite{StreetetAl02APL}, and divide the total source-drain
voltage $V_{\rm D}$ into a channel component and a voltage dropped at
the contacts, $V_{\rm C}$ .  We assume, as supported by scanning
potentiometry\cite{BurgietAl03JAP}, that $V_{\rm C}$ is dominantly
dropped at the injecting contact for metals with a significant
$\Delta$.  We use $I_{\rm D}-V_{\rm D}$ data from a given series of
devices of varying channel length, $L$, and fixed width, $W$, to
extract both $\mu$ and $I_{\rm D}-V_{\rm C}$ for this OSC/metal
interface.  As expected, the $I_{\rm D}-V_{\rm C}$ characteristics of
a specific OSC/metal interface are unique at a given temperature and
gate voltage, independent of $L$.  We analyze the field and
temperature dependence of the injected current through a recent
analytical model\cite{ArkhipovetAl98JAP} of charge injection from a
metallic electrode into a random hopping system.  With reasonable fit
parameters, this model agrees well with the observed temperature and
field dependence of the injected current.  We also discuss the
distance scale over which $V_{\rm C}$ is dropped, and further
experimental avenues to explore.

\section*{II. EXPERIMENTAL DETAILS}

Devices are made in a bottom contact
configuration\cite{HamadanietAl04APL} on a degenerately doped $p+$
silicon substrate used as a gate. The gate dielectric is 200~nm of
thermal SiO$_{2}$.  Source and drain electrodes are patterned using
electron beam lithography in the form of an interdigitated set of
electrodes with a systematic increase in the distance between each
pair.  The channel width, $W$, is kept fixed for all devices.  Three
different kinds of metallic electrodes (Au, Cr, Cu) were then
deposited by electron beam evaporation followed by lift off. (25~nm of
each, preceded by 2.5~nm of Ti adhesion layer; no Ti layer for Cr
samples). Au electrodes were cleaned for one minute in a 1:1 solution
of NH$_{4}$OH: H$_{2}$O$_{2}$ (30\%), rinsed in de-ionized water, and
exposed for about 1 min to oxygen plasma.  The Cr samples were cleaned
in the same manner followed by a last step dipping in a buffered HF
solution for under 10 seconds. The HF is believed to etch the native
SiO$_{2}$ oxide, exposing a fresh layer of dielectric. Cu electrodes
on the other hand were only exposed to less than 25 seconds of O$_{2}$
plasma to clean the organic residue from the lift off. We found that
Cu samples exposed to any cleaning procedure except for short O$_{2}$
plasma generally exhibited very poor transport properties.

The organic semiconductor is 98\% regio-regular P3HT\cite{purchase} a
well studied
material\cite{BaoetAl96APL,SirringhausetAl98Science,SirringhausetAl99Nature}.
As received, RR-P3HT is dissolved in chloroform at a 0.02\% weight
concentration, passed through PTFA 0.02 micrometer filters, and
solution cast onto the clean substrates, with the solvent allowed to
evaporate in ambient conditions. The resulting films are tens of
nanometers thick as determined by atomic force microscopy. The
measurements are performed in vacuum ($\sim 10^{-6}$~Torr) in a
variable-temperature probe station using a semiconductor parameter
analyzer (HP4145B).

\section*{III. RESULTS AND DISCUSSION}

The devices operate as standard $p$-type FETs in accumulation
mode\cite{AleshinetAl01SM,MeijeretAl02APL}.  With the source electrode
grounded, the devices are measured in the shallow channel regime
($V_{\rm D}< V_{\rm G}$). Here we mainly concentrate on the
experimental results of charge injection from Cr and Cu electrodes, as
results on Au samples have been published
elsewhere\cite{HamadanietAl04APL,HamadanietAl04JAP}.

\subsection*{A.  Extracting contact current-voltage characteristics}

\begin{figure}[h!]
\begin{center}
\includegraphics[clip, width=7.5cm]{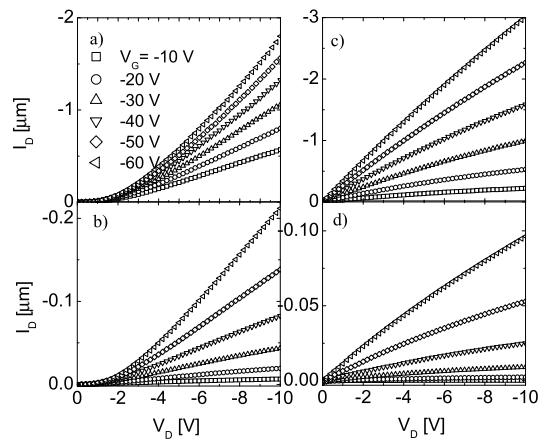}
\end{center}
\vspace{-3mm}
\caption{\small (a) Transport characteristics ($I_{\rm D}-V_{\rm D}$) of a P3HT OFET with Cr source and drain electrodes, with $L = 25~\mu$m, $W=200~\mu$m at 300~K for several gate voltages. (b) Same device at 160~K.  (c) Analogous plot for a sample with Au electrodes of the same geometry, at 300~K and (d) at 160~K.}
\label{fig:fig1}
\end{figure}

Figures~\ref{fig:fig1}a and \ref{fig:fig1}b show the transport
characteristics ($I_{\rm D}-V_{\rm D}$) of a Cr device with $L =
25$~$\mu$m and $W = 200$~$\mu$m at $T = 300$~K and $T = 160$~K for a
series of $V_{\rm G}$s. The rise of current with voltage in the low
$V_{\rm D}$ regime ($V_{\rm D} < 3$~V) is slightly super-quadratic. At
higher drain voltages, the current becomes less nonlinear, at least in
part due to saturation effects in the transistor.  Our focus here in
on low source-drain bias voltages. For comparison, $I_{\rm D}-V_{\rm
D}$ plots for an Au sample with the same geometric parameters as above
is also included in Figs.~\ref{fig:fig1}c and \ref{fig:fig1}d.  

The severe nonlinearity observed in the Cr data, in contrast to the
linear injection from Au, is traditionally attributed to the existence
of a Schottky energy barrier, $\Delta$, between the Fermi level of the
metal and the HOMO of the polymer. It is generally
accepted\cite{ChiguvareetAl03JAP,ArkhipovetAl98JAP} that the size of
$\Delta$ alone determines the nature of transport in a device, {\it
i.e.}, injection limited vs. bulk (or space charge limited) transport.
In the space-charge case, most relevant when $\Delta$ is relatively
small, the easily injected carriers significantly alter the local
electric field from the average field imposed by the electrode
potentials, and correspondingly limit the current.  From individual
$I_{\rm D}-V_{\rm D}$ traces, it can be difficult to distinguish
between space charge effects and nonlinear contact injection because
space charge limited transport is also significantly
nonlinear\cite{Murgatroyd70JPD}.  One means of distinguishing the two
is the detailed dependence of current on $V$ and
$L$\cite{StreetetAl02APL,ArkhipovetAl98JAP,HamadanietAl04JAP}.  A weak
temperature dependence of current is also another signature of
injection limited transport from metal into organic
semiconductor\cite{ReynaertetAl04APL}.

The injection properties of the contact can be
examined\cite{StreetetAl02APL} by splitting the channel into two
regimes of contact and the main channel. A voltage of $V_{\rm C}$ is
dropped across the contact with the remaining $V_{\rm ch} = V_{\rm
D}-V_{\rm C}$ across the main channel. Using the charge control
model\cite{Shurbook}, $I_{\rm D}$ can be written as:
\begin{equation}
I_{\rm D} = WC \mu [V_{\rm G}-V_{\rm T}-V(x)]\frac{dV}{dx},
\label{eq:eq1}
\end{equation}
where $V(x)$ is the potential in the channel at some position $x$, $V_{\rm
T}$ is the threshold voltage, $C$ is the capacitance per unit area of
the gate dielectric, and $\mu$ is the intrinsic channel
mobility. Integration of Eq.~(\ref{eq:eq1}) from $x = 0$ to $L-d$
gives:
\begin{equation}
\frac{I_{\rm D}}{WC \mu}(L-d)=(V_{\rm G}-V_{\rm T})(V_{\rm D}-V_{\rm C})-\frac{1}{2}(V_{\rm D}^{2}-V_{\rm C}^{2}),
\label{eq:eq2}
\end{equation}
where $V_{\rm C}$ is dropped across $d$, a characteristic depletion
length near the contacts. In this treatment, we assume $V_{\rm C}$ to
be entirely dropped across the injecting contact.  Scanning
potentiometry experiments in this
material\cite{SeshadrietAl01APL,BurgietAl02APL,BurgietAl03JAP} have
previously shown that, in systems with significant $\Delta$, most of
the potential drop due to contacts occurs at the source, where holes
are injected into the channel.

Eq.~(\ref{eq:eq2})can be used to extract a value of $V_{\rm C}$ for
any pair of $(V_{\rm D}, I_{\rm D})$ data, though there is no
independent way of knowing the correct value of $\mu$.  With an array
of devices, one can use the length dependence of $I_{\rm D}$ to
address this difficulty.  At a given $T$ and $V_{\rm G}$, a series of
$I_{\rm D}-V_{\rm D}$ data is collected from devices with different
channel lengths. The corresponding $I_{\rm D}-V_{\rm C}$ is calculated
from Eq.~(\ref{eq:eq2}) for all the different $L$.  If the contact and
channel transport properties in each device are identical, the correct
value of $\mu$ would make all the different $I_{\rm D}-V_{\rm C}$
curves collapse onto one, since the injection characteristics of a
particular OSC/metal interface should be unique and set by material
properties and the (fixed) channel width and electrode geometry.  This
technique allows for the simultaneous extraction of $\mu$ and $I_{\rm
D}-V_{\rm C}$. Since the average source-drain field in our devices is
low ($< 10^3$~V/cm), no significant field dependence of $\mu$ is
expected\cite{StreetetAl02APL} or observed.

\begin{figure}[h!]
\begin{center}
\includegraphics[clip, width=7.5cm]{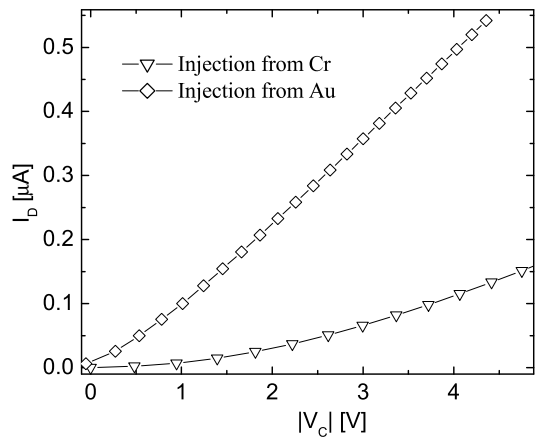}
\end{center}
\vspace{-3mm}
\caption{\small Extracted $I_{\rm D}-V_{\rm C}$ for a series of
devices of width 200~$\mu$m with alternating Cr and Au electrodes at
240~K and $V_{\rm G}=-80$~V.  Upper curve shows injection of holes
from Au, while lower curve shows injection from Cr.  Injection from Au
is more linear and allows higher currents at lower voltages.  The Au
data have been shifted to lower $|V_{\rm C}|$ by 0.5~V to account for
a small contact voltage at the drain.}
\label{fig:fig2}
\end{figure}

To confirm this method of extracting $\mu$ and $I_{\rm D}-V_{\rm C}$,
we fabricated a series of devices (in a two-step lithography process)
with {\it alternating} Au and Cr electrodes.  The data is then taken
twice for each device, once with the source electrode on Cr with the
drain on Au and the second time {\it vice versa}.  Fig.~\ref{fig:fig2}
shows a plot of extracted $I_{\rm D}-V_{\rm C}$ for injection from Cr
and Au at $T = 240$~K and $V_{\rm G}=-80$~V.  We noticed that there is
still a minute nonlinearity present in data for Au at that is not
present in all-Au devices.  We believe that this is consistent with a
small contact voltage at the drain, as was seen in the potentiometry
profile of Cr/P3HT devices in Ref.~\cite{BurgietAl03JAP}.  The Au data
in Fig.~\ref{fig:fig2} have been shifted toward lower $|V_{\rm C}|$ by
0.5~V to account for this.  The value of $\mu$ that collapses the
different length-dependent data for injection from Cr onto one $I_{\rm
D}-V_{\rm C}$ curve is {\it identical} to that inferred from the
length-dependence of the channel resistance when injection is from Au
in the same devices.  This demonstrates that this procedure of
extracting $I_{\rm D}-V_{\rm C}$ is well-founded.

The mobilities in the Au/Cr devices are lower than those seen in all
Au or all Cr source/drain samples (discussed below). We believe this
to be due to inferior surface cleanliness of samples made in the
two-step lithography technique.  The contact resistance data for
injection from Au agree {\it quantitatively} with the data observed in
previous Au samples\cite{HamadanietAl04APL}.

\begin{figure}[h!]
\begin{center}
\includegraphics[clip, width=7.5cm]{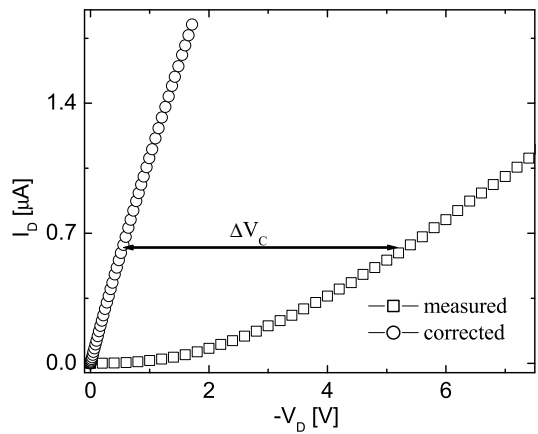}
\end{center}
\vspace{-3mm}
\caption{\small Measured $I_{\rm D}-V_{\rm D}$, and the same data
corrected for contact voltages ({\it i.e.} $I_{\rm D}-V_{\rm ch}$) for
the Cr electrode device shown in Fig.~\ref{fig:fig1}, at 290~K and
$V_{\rm G}=-60$~V.  The device is clearly quite contact limited.}
\label{fig:fig3}
\end{figure}

Fig.~\ref{fig:fig3} shows a plot of measured $I_{\rm D}-V_{\rm D}$ and
the current corrected for contact voltages, {\it i.e.} $I_{\rm
D}-V_{\rm ch}$, for the all-Cr sample described above at $V_{\rm G} =
-60$~V. As seen from the plot, most of the total voltage is dropped
across the contact, making these devices severely contact limited. For
example, for a drain voltage of 2~V, $V_{\rm C}/V_{\rm ch}\sim 30$.

\begin{figure}[h!]
\begin{center}
\includegraphics[clip, width=7.5cm]{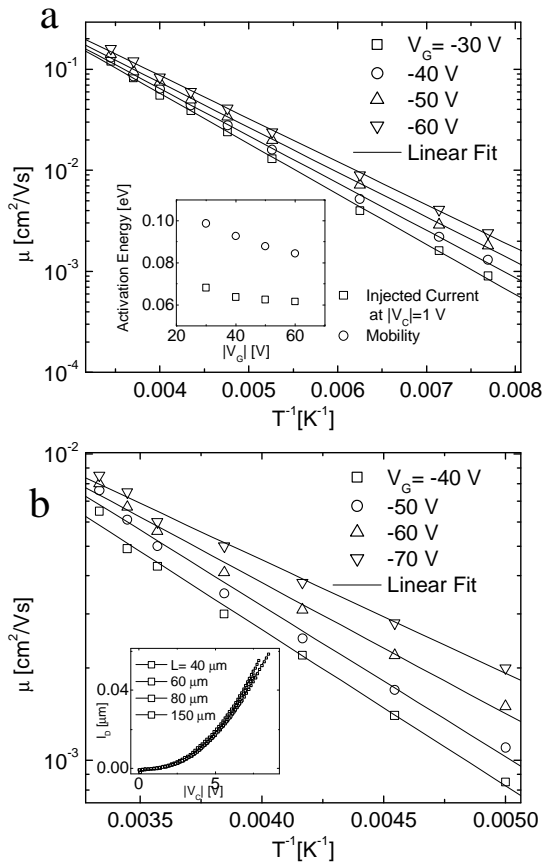}
\end{center}
\vspace{-3mm}
\caption{\small Temperature dependence of the extracted channel
mobility as a function of $T^{-1}$ for sets of devices with (a) Cr and
(b) Cu source and drain electrodes at several gate voltages.  Inset in
(a): Activation energies of the injected current and $E_{a}$ as a
function of $V_{\rm G}$.  Inset in (b): An example of ``collapsed''
$I_{\rm D}-V_{\rm C}$ data for devices of different values of $L$ with
Cu electrodes, $W = 400~\mu$m, 240~K, $V_{\rm G}=-70$~V.}
\label{fig:fig4}
\end{figure}

Fig.~\ref{fig:fig4}a shows the temperature dependence of $\mu$
extracted this way as a function of $T^{-1}$ for a set of devices with
all-Cr source/drain electrodes.  The temperature dependence is well
approximated as thermal activation consistent with simple hopping of
carriers between localized states in the channel.  The activation
energies $E_{\rm a}$ for the mobility are quantitatively similar to
those seen in all-Au devices\cite{HamadanietAl04APL}.  The inset in
Fig.~\ref{fig:fig4}a shows that the activation energies of the
injected current (at $V_{\rm C}=1$~V) are {\it smaller} than $E_{\rm
a}$.  In agreement with others'
results\cite{BurgietAl03JAP,WoudenberghetAl01APL}, this is
{\it inconsistent} with the simple thermionic diffusion model of injection.
As discussed in the next section, the hopping injection model predicts
this weak temperature dependence.

A similar plot for the temperature dependence of channel mobility in a
sample with Cu source/drain electrodes is shown in
Fig.~\ref{fig:fig4}b.  The values of $\mu$ in Cu devices are
consistently lower than those seen in Cr or Au, though the activation
energies are very similar.  The reason for these lower mobilities
remains unclear, since contact effects have been accounted for.
However, it is commonly observed\cite{BurgietAl03JAP,NelsonetAl98APL}
that the field effect mobility can be significantly different in
nominally identically prepared samples.  The values of contact
resistivity, $R_{\rm c}W$ are also higher in Cu devices. At a small
$V_{\rm C}\sim$~1~V, $R_{\rm c}W[{\rm Cu}] \sim$3300~k$\Omega$-cm,
while the corresponding value for Cr is $R_{\rm c}W[{\rm
Cr}]\sim$125~k$\Omega$-cm.  For devices with Au source/drain
electrodes\cite{HamadanietAl04APL}, $R_{\rm c}W\sim$~10~k$\Omega$-cm
was obtained.  The increase in contact resistivity from Au to Cu is
consistent with the increase in value of $\Delta$ as described below.
The inset in Fig.~\ref{fig:fig4}b shows an example of a coalesced plot
of $I_{\rm D}-V_{\rm C}$ for different values of $L$ for the Cu sample
at $T = 240$~K and $V_{\rm G}=-70$~V. A single mobility value of $3.8
\times 10^{-3}$~cm$^{2}$/Vs was used to obtain this collapse.

\subsection*{B.  Interpretation and modeling}

Having extracted nonlinear current-voltage characteristics for the
injecting contacts, we analyze the resulting data in terms of a
particular model of injection into disordered polymer semiconductors.
As mentioned earlier, the diffusion thermionic models are
insufficient to account for the weak temperature dependence of the
observed injected current. The analytic treatment by Arkhipov {\it et
al.}\cite{ArkhipovetAl98JAP} of charge injection from a metallic
electrode into a random hopping system has been
shown\cite{WoudenberghetAl01APL,ReynaertetAl04APL} to consistently
explain the field and temperature dependence of charge injection in
organic Schottky diode-like structures.  In this picture, the weak
temperature dependence of the injection current is a consequence of a
Gaussian distribution of states\cite{ArkhipovetAl99PRB}. Therefore the
injection process is easier at lower temperatures, leading to a weaker
temperature dependence of the current.  Here we apply Arkhipov results
to our charge injection data in OFETs and show that this treatment,
with reasonable parameters, is consistent with the measured data.

Key to this analysis is the
conclusion\cite{BurgietAl03JAP,LietAl03JAP} that a small depletion
region forms in the vicinity of the contacts, and that $V_{\rm C}$ is
dropped across this region at the injecting contact.  We note that the
values of $V_{\rm C}$ can be as large as a few Volts.  Establishing
the distance scale, $d$, relevant to converting this into the electric
field at the contact is nontrivial, though reasonable bounds may be
placed on this parameter.  The resolution of existing scanning
potentiometry data in similar OFET
structures\cite{BurgietAl02APL,BurgietAl03JAP} establishes that $d$
cannot exceed $\sim$~400~nm.  Furthermore, the lack of breakdown or
irreversible device damage implies that the injecting field must be
below the breakdown field of the OSC, so that $d$ must be larger than
$\sim$10~nm.  After presenting the analysis of the $I_{\rm D}-V_{\rm
C}$ data, we return to this issue below.

In this 1d model\cite{ArkhipovetAl98JAP}, the transport of carriers
takes place in a hopping system of Gaussian energy distribution in
close contact with the metallic electrode. This density of states
(DOS) is given by:
\begin{equation}
g(E)=\frac{N_{\rm t}}{\sqrt{2 \pi}\sigma} \exp \left(-\frac{E^{2}}{2 \sigma^{2}}\right),
\label{eq:eq3}
\end{equation}
where $N_{\rm t}$ is the total spatial density of localized states,
with $\sigma$ as the variance of the gaussian distribution centered
about $E = 0$.  The emphasis is placed on the primary injection event
where a carrier from the metal is injected into a localized state a
distance $x_{0} > a$ from the interface, where $a$ is the intersite
hopping distance. The potential of this carrier at any distance $x$
from the interface is given by
\begin{equation}
U(x,E)=\Delta - \frac{e^{2}}{16 \pi \epsilon_{0}\epsilon x} - eF_{0}x + E,
\label{eq:eq4}
\end{equation}
where $\Delta$ is the energy difference between the Fermi level of the
metal and the center of DOS in the semiconductor, $F_{0}$ is the
external field at the contact, $e$ is the elementary charge, and
$\epsilon$ is the relative dielectric constant of the polymer. Once a
carrier is injected into a localized state in the polymer, it can
either go back to the metal due to the attractive image potential, or
escape with a finite probability to diffuse into the bulk. The escape
probability can be solved using the 1d Onsager problem as outlined in
detail in Ref.~\cite{ArkhipovetAl98JAP}. The final result predicts the
injection current density as follows:
\begin{eqnarray}
J_{\rm inj}& = & e \nu \left( \int_{a}^{\infty} dx \exp \left[- \frac{e}{k_{\rm B}T}\left(F_{0}x + \frac{e}{16 \pi \epsilon_{0} \epsilon x}\right)\right]\right)^{-1} \times \nonumber \\
& & \hspace{-7mm} \int_{a}^{\infty} dx_{0} \exp(2 \gamma x_{0}) \int_{a}^{x_{0}} dx \exp \left[- \frac{e}{k_{\rm B}T}\left(F_{0}x + \frac{e}{16 \pi \epsilon_{0} \epsilon x}\right)\right] \times \nonumber \\
& & \hspace{-7mm} \int_{-\infty}^{\infty}dE' Bol(E') g[U(x_{0})-E'].
\label{eq:eq5}
\end{eqnarray}
Here, $\nu$ is the attempt-to-jump frequency, $T$ is the temperature, $\gamma$ is inverse
localization length, and the Boltzmann function $Bol(E)$ is defined as:
\begin{eqnarray}
Bol(E)& = & \exp(-E/k_{\rm B}T),  E>0,\nonumber \\
& = & 1,  E < 0.
\label{eq:eq6}
\end{eqnarray}

To apply this model, we first need to fix
the parameters $\sigma$, $a$, and $\gamma$. It is possible to extract
$\sigma$ from a model of carrier transport in a disordered Gaussian
density of states\cite{MartensetAl00PRB,Bassler93PSSB} by plotting
$\ln(\mu_{0})$ vs. $T^{-2}$, where $\mu_{0}$ is the value of the
zero-field mobility.  We note that our data appear to be better
described as exponential in $T^{-1}$ rather than $T^{-2}$; nonetheless
this procedure provides an estimate for a value of $\sigma$.
Calculation of the values of $a$ and $\gamma$ \cite{MartensetAl00PRB}
can be difficult, as one has to use the strong field dependence of
mobility.  As mentioned earlier, our data are acquired in a low enough
source-drain average field that no field-dependence of $\mu$ may be
inferred.  Therefore, we chose $a$ and $\gamma$ consistent with
reported values in literature\cite{MartensetAl00PRB} or previous
experiments\cite{HamadanietAl04JAP}.  We note that changing $a$ or
$\gamma$ over a reasonable range mainly affects the overall prefactor
of the current (as described below), without significantly altering
the shapes of the predicted curves.

With $\sigma$, $a$, and $\gamma$ held fixed, the only parameters that
can be adjusted to fit Eq.~(\ref{eq:eq5}) to a plot of data are
$\Delta$, a prefactor $K \equiv A \nu N_{\rm T}$ (where $A$ is an
effective injection area), and $d$, where $F_{0}=V_{\rm C}/d$.  We
observed, as discussed in detail in Ref.~\cite{ArkhipovetAl98JAP}, that
the nonlinearity in a plot of $I_{\rm D}-V_{\rm C}$ is mainly
controlled by the value of $\Delta$ and the strength of the electric
field.  At $F_{0} \sim 5\times 10^{7}$~V/m or higher, the plots are
severely nonlinear and the temperature dependence of the current would
be extremely weak.  At lower fields, the nonlinearity is less severe
and the temperature dependence is stronger.  Therefore, the value of
$d$ is paramount, and constrained as described above.

\begin{figure}[h!]
\begin{center}
\includegraphics[clip, width=7.5cm]{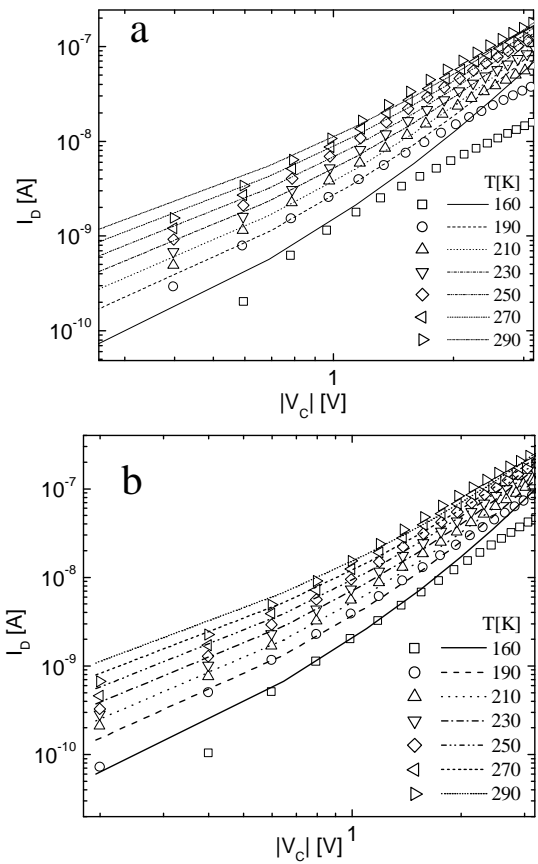}
\end{center}
\vspace{-3mm}
\caption{\small $I_{\rm D}-V_{\rm C}$ data from a set of Cr electrode devices, together with curves from the model of Eq.~(\ref{eq:eq5}) at gate voltages (a) -30~V, and (b) -60~V.}
\label{fig:fig5}
\end{figure}

\begin{figure}[h!]
\begin{center}
\includegraphics[clip, width=7.5cm]{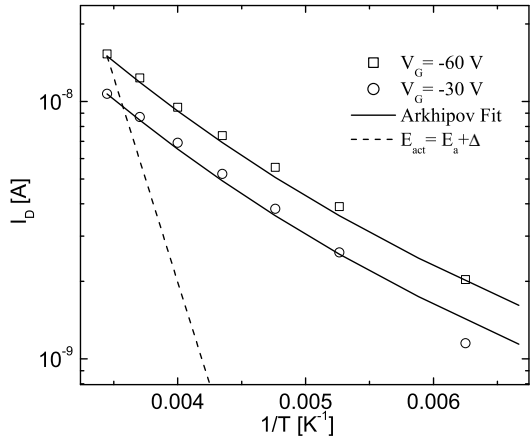}
\end{center}
\vspace{-3mm}
\caption{\small Temperature dependence of the injected current for Cr
electrodes at low $V_{\rm C}$, with Eq.~(\ref{eq:eq5}) fits to the
data.  The dashed line shows the predicted $T$ dependence of the
thermionic diffusion model for the same value of $\Delta$.}
\label{fig:fig6}
\end{figure}

\begin{figure}[h!]
\begin{center}
\includegraphics[clip, width=7.5cm]{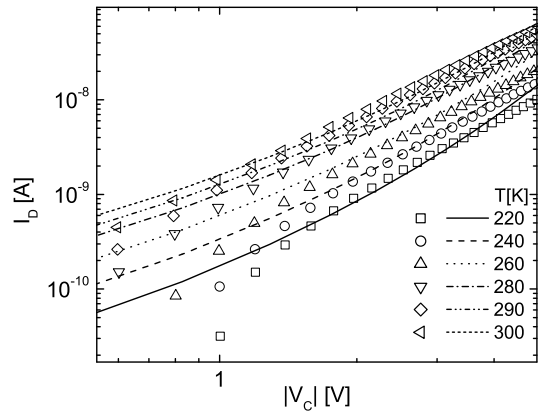}
\end{center}
\vspace{-3mm}
\caption{\small $I_{\rm D}-V_{\rm C}$ for devices with Cu electrodes
($W = 400$~nm) at $V_{\rm G}=-60$~V over a representative temperature
range, together with results from Eq.~(\ref{eq:eq5}) and the
parameters of Table~\ref{table}.}
\label{fig:fig7}
\end{figure}

Having expected $\Delta \sim 0.3$~eV, we find that $d$ cannot be below
100 nm within this model; the resulting large values of injecting
field $F_{0}$ would yield temperature and field dependences
inconsistent with those observed.  Table~\ref{table} summarizes the
parameters used to model the injection data for both all-Cr and all-Cu
sets of devices.  Figures~\ref{fig:fig5}a and b show plots of $I_{\rm
D}-V_{\rm C}$ and the corresponding numerical integration of
Eq.~(\ref{eq:eq5}) using parameters given in Table~\ref{table} (with
an appropriate value of prefactor $K$) for injection from Cr over a
representative range of temperature for gate voltages -30~V and -60~V.
Fig.~\ref{fig:fig6} shows the temperature dependence of the injected
current in low $V_{\rm C}$ regime and the Arkhipov fit to the data.
Notice that the predicted temperature dependence in the diffusion
thermionic model is much stronger than the Arkhipov model if the same
$\Delta = 0.23$~eV is used.  Fig.~\ref{fig:fig7} shows a plot of
$I_{\rm D}-V_{\rm C}$ for a Cu sample at $V_{\rm G}=-60$~V. The fit to
the data is valid only in the low $V_{\rm C}$ regime as saturation
effects in the transistor start to affect $I_{\rm D}-V_{\rm D}$ data
at large $V_{\rm D}$.  Note that the procedure outlined above to
extract the $I_{\rm D}-V_{\rm C}$ data assumes that devices are firmly
in the gradual channel limit, with no saturation effects.  Also, the
effects of leakage currents to the gate electrode in the immediate
vicinity of $V_{\rm C} = 0$ at low temperatures can be seen in
Figs.~\ref{fig:fig5} and \ref{fig:fig7}.

$N_{\rm t}$ is the only gate-dependent parameter in this model.  The
parameters listed in Table~\ref{table} are kept fixed for all
temperatures and all gate voltages.  Since the prefactor in
Eq.~(\ref{eq:eq5}) is the product $A \nu N_{\rm t}$, it is therefore
not possible to determine an exact value for just $N_{\rm t}$.  For a
cross-sectional area of injection of $A \sim$~25~nm$\times 2 \times
10^{5}$~nm and $\nu \sim 10^{13}$~s$^{-1}$, we find $N_{\rm t} \sim
1.1 \times 10^{22}$~cm$^{-3}$ for $V_{\rm G} = -60$~V and $N_{\rm
t}\sim 8.3 \times 10^{21}$~cm$^{-3}$ for $V_{\rm G} = -30$~V.  These
values are consistent with other
experiments\cite{WoudenberghetAl01APL,ReynaertetAl04APL}.

\begin{table}
\caption{\small Parameters used to model the $I_{\rm D}-V_{\rm C}$
data of this study within the charge injection treatment of
Ref.~\protect{\cite{ArkhipovetAl98JAP}} for all $T$.  The relative
dielectric constant $\epsilon$ of the polymer was assumed to be 3.  At
each gate voltage a single numerical prefactor was the only necessary
adjustment.}
\vspace{4mm}
\begin{tabular}{|c c c c c c|}
\hline
Contact  & ~~$\sigma$~~ & ~~$a$~~ & ~~$\gamma$~~  & ~~$\Delta$~~ & ~~$d$~~ \\
metal & ~[eV]~ & ~[nm]~ & ~[nm$^{-1}]$~ & ~[eV]~ & ~[nm]~\\
\hline
Cr  & 0.046 & 1.6 & 4.35 & 0.23 & 150 \\
Cu  & 0.046 & 1.6 & 4.35 & 0.31 & 230 \\
\hline
\end{tabular}
\label{table}
\vspace{-3mm}
\end{table}

Table~\ref{table} shows that the obtained injection barrier height for
copper is about 80~meV higher than that for Cr.  This difference in
the barrier energy is not unreasonable, and may be
attributed\cite{BurgietAl03JAP} to an interfacial dipole layer at the
interface changing $\Delta$ by a small amount.  {\it In-situ}
ultraviolet photoemission spectroscopy measurements would be
well-suited to testing this
hypothesis\cite{HilletAl98APL,CrispinetAl02JACS,KochetAl03APL}.  The
higher injection barrier for copper is consistent with the observed
higher contact resistivity and lower overall currents observed in Cu.

The length of the presumed depletion region is also a bit higher in Cu
samples (by $\sim$~80~nm), though it does not necessarily reveal why
the mobility is lower in these devices.  The origin of these depletion
regions in the vicinity of the contacts is not understood in detail.
Recent 2d electrostatic modeling\cite{LietAl03JAP} of OFETs has shown
that the effect of significant energy barriers at the injecting
electrode is formation of regions of low carrier concentration (and
mobility) near the contacts. These studies place the extent of these
regions at about 100~nm from the contact, depending on $V_{\rm G}$.
Another possible origin for regions of reduced mobility near metal
contacts with significant barriers is charge transfer and band bending
near the interface.  Since conduction in these materials can be
treated as percolative variable range
hopping\cite{VissenbergetAl98PRB}, $\mu$ is a natural function of the
density of available hopping sites.  The occupation of those sites can
be strongly modified by interfacial charge transfer between the metal
and the OSC.  Improved local probes (nm-resolution scanning
potentiometry, cross-sectional scanning tunneling microscopy) would be
extremely useful in better understanding these depletion regions.

\section*{V. CONCLUSIONS}

Transport properties of a series of organic field effect transistors
with P3HT as the active polymer layer and Cr, Cu and Au as the
source/drain electrodes were examined over a temperature range.  The
contact current-voltage characteristics for these devices were
extracted from the length dependence of conductance, with the
assumption that the injection barrier primarily applies to holes being
injected from the source.  This procedure was checked for consistency
using devices with electrodes of alternating metal composition.  The
data confirm that the weak temperature dependence of the injected
current cannot be simply explained using the general
diffusion-thermionic emission models.  With reasonable values of
parameters, a more sophisticated model of hopping injection into a
disordered density of localized states, with emphasis on the primary
injection event, is consistent with the field and the temperature
dependence of the data over a broad range of temperatures and gate
voltages.

\section*{VI. ACKNOWLEDGMENT}

The authors gratefully acknowledge the support of the Robert A. Welch
Foundation and the Research Corporation.




\end{document}